\documentclass{emulateapj}

\newcommand{\sgra}{Sgr~A*}

\newcommand{\rmm}{rad~m$^{-2}$}
\newcommand{\rin}{$r_\mathrm{in}$}

\newcommand{\msunyr}{$M_\sun$~yr$^{-1}$}

\defcitealias{MarroneE06}{M06}

\shorttitle{Faraday Rotation in Sagittarius~A*}
\shortauthors{Marrone et al.}

\begin{document}

\title{An Unambiguous Detection of Faraday Rotation in Sagittarius A*}
\author{Daniel~P.~Marrone\altaffilmark{1,2},
James~M.~Moran\altaffilmark{1}, Jun-Hui~Zhao\altaffilmark{1}, and
Ramprasad Rao\altaffilmark{3}}
\altaffiltext{1}{Harvard-Smithsonian Center for Astrophysics, 
60 Garden Street, Cambridge, MA 02138, {\it jmoran@cfa.harvard.edu},
{\it jzhao@cfa.harvard.edu}}
\altaffiltext{2}{Current Address: Jansky Fellow, NRAO, University of
Chicago, 5640 South Ellis Avenue, Chicago, IL 60637, {\it
dmarrone@oddjob.uchicago.edu}}
\altaffiltext{3}{Institute of Astronomy and Astrophysics, Academia
Sinica, P.O. Box 23-141, Taipei 10617, Taiwan., {\it
rrao@sma.hawaii.edu}}

\slugcomment{To appear in the Astrophysical Journal Letters}

\begin{abstract}
The millimeter/submillimeter wavelength polarization of \sgra\ is
known to be variable in both magnitude and position angle on time
scales down to a few hours. The unstable polarization has prevented
measurements made at different frequencies and different epochs from
yielding convincing measurements of Faraday rotation in this
source. Here we present observations made with the Submillimeter Array
polarimeter at 227 and 343~GHz with sufficient sensitivity to
determine the rotation measure at each band without comparing position
angles measured at separate epochs. We find the 10-epoch mean rotation
measure to be $(-5.6\pm0.7)\times10^5$~\rmm; the measurements are
consistent with a constant value. We conservatively assign a 3$\sigma$
upper limit of $2\times10^5$~\rmm\ to rotation measure changes, which
limits accretion rate fluctuations to 25\%. This rotation measure
detection limits the accretion rate to less than
$2\times10^{-7}$~\msunyr\ if the magnetic field is near equipartition,
ordered, and largely radial, while a lower limit of
$2\times10^{-9}$~\msunyr\ holds even for a sub-equipartition,
disordered, or toroidal field. The mean intrinsic position angle is
167\arcdeg$\pm$7\arcdeg\ and we detect variations of $31^{+18}_{-9}$
degrees. These variations must originate in the submillimeter
photosphere, rather than arising from rotation measure changes.
\end{abstract}

\section{Introduction}
The linear polarization of \sgra\ was first detected by
\citet{AitkenE00} above 100~GHz, after unsuccessful searches at
lower frequencies
\citep[e.g.][]{BowerE99-lp1,BowerE99-lp2}. Subsequent interferometer
observations have shown the polarization to vary in position angle
\citep{BowerE05} and fraction (\citealt{MarroneE06}; hereafter M06),
with variability occurring on timescales comparable to those of the
previously observed total intensity variations
\citepalias{MarroneE06}. The variability may be intrinsic to the
source or due to propagation effects, but the short timescales suggest
processes very close to the black hole are responsible and thus
polarization should be a useful tool for the study of \sgra.

One aspect of the polarization that has yet to be exploited is its
variation with frequency, particularly the frequency-dependent
orientation change known as Faraday rotation. The presence of
polarization was used immediately after its detection to argue that
the infalling plasma must be tenuous and the mass accretion rate low
(less than $10^{-6}$~\msunyr), because larger accretion rates would
depolarize the emission through extreme Faraday rotation gradients
\citep{QuatGruz00-LP,Agol00}. Constraints have also been
derived through careful examination of the spectrum
\citep[e.g.][]{MeliaLiuCoker00,MeliaLiuCoker01}. However, further
progress on determining the accretion rate and its variability has
awaited measurement of the Faraday rotation measure (RM), a difficult
task because of significant variability of \sgra\ and the diminished
effect of rotation at high frequencies. To date there have been three
RM determinations from non-simultaneous observations
(\citealt{BowerE03}; \citetalias{MarroneE06}; \citealt{MacquartE06}),
but none of them is robust. Most recently, \citet{MacquartE06}
estimated the RM to be $-4.4\times10^5$~\rmm\ from their 83~GHz
polarization data and all previous data. However, their analysis
allows a much lower RM due to the 180\arcdeg\ degeneracy of the
polarization position angle (see
\S~\ref{s-RMX}).

Simultaneous measurements are a much more secure way to determine the
RM because they are insensitive to source variability. Such data can
be used to measure the RM, the intrinsic polarization direction, and
the variability of each. Isolating these changes is crucial to
understanding the conditions in the inner accretion flow; the RM can
be used to place upper and lower limits on the accretion rate, while
variations in the polarization and RM with frequency and time can be
used to examine the structure of the flow. The instantaneous frequency
coverage of available instruments has not yet been adequate to show
Faraday rotation in a single epoch, with the lowest upper limit
($|$RM$|\le$$7\times10^5$~\rmm) from early Submillimeter
Array\footnote{The Submillimeter Array is a joint project between the
Smithsonian Astrophysical Observatory and the Academia Sinica
Institute of Astronomy and Astrophysics, and is funded by the
Smithsonian Institution and the Academia Sinica.} (SMA) 345~GHz
polarimetric observations \citepalias{MarroneE06}. Here we report SMA
measurements of the RM and intrinsic polarization changes of \sgra,
the first observations to detect the RM with statistical significance
and the only measurement made without resorting to comparisons between
position angles measured at different epochs. The observations and
calibration procedures are detailed in \S~\ref{s-obs}, the RM and the
variability of the RM and intrinsic polarization are examined in
\S~\ref{s-RMX}, and we discuss the implications for the accretion rate
and properties of \sgra\ in \S~\ref{s-disc}.

\section{Observations}
\label{s-obs}
The SMA made polarimetric observations of \sgra\ in 2005 June and
July. In the ten nights without weather or technical problems there
are six (four) tracks in the 230~GHz (345~GHz) band. The 230~GHz
observations used a local oscillator (LO) frequency of 226.9~GHz,
providing simultaneous measurements of sidebands centered on 221.9 and
231.9~GHz, except for one observation made at LO=225.55~GHz. The
higher frequency data use an LO frequency of 343~GHz. The weather was
excellent in the usable nights, with a typical zenith opacity at
225~GHz of $0.06-0.08$, or $0.19-0.27$ when scaled to 343~GHz. All
data were obtained with five to seven antennas in the compact
configuration ($7-70$~meter baselines), resulting in a synthesized
beam width of 2\farcs0$\times$4\farcs0 at 230~GHz and
1\farcs6$\times$3\farcs2 at 345~GHz on \sgra. For the purposes of
self-calibration and polarization extraction, baselines shorter than
$20 k\lambda$ were removed to exclude the extended emission around
\sgra. Gain calibrators included J1743-038, J1733-130, J1924-292, and
Ceres. Very similar polarization results were obtained when the
calibrator gains were applied and when \sgra\ was self-calibrated. The
flux density scale was derived from observations of planets and their
moons, with an expected accuracy of better than 10\%. Flux densities
in each Stokes parameter were obtained through point-source fits to
the visibilities. The observed properties of \sgra\ in all ten epochs
are listed in Table~\ref{t-polObs}.

\begin{deluxetable*}{ccccccccc}
\tabletypesize{\scriptsize}
\setlength{\tabcolsep}{0.06in}
\tablecolumns{9}
\tablewidth{0pt}
\tablecaption{Polarization and Rotation Measure of Sagittarius~A*\label{t-polObs}}
\tablehead{Date & $\nu$ & $I$\tablenotemark{a} & $Q$\tablenotemark{a} 
	& $U$\tablenotemark{a} & $m$ & $\chi$ 
	& RM & $\chi_0$ \\
 & (GHz) & (Jy) & (mJy) & (mJy) & (\%) & (deg) &
	($10^5$ rad/m$^2$) & (deg)}
\startdata
2005 Jun 4:\hfill\hfill & \nodata & \nodata & \nodata & \nodata & \nodata & \nodata & $-6.7\pm2.9$ & $ 215\pm  29$ \\ 
\hfill USB \dotfill & 230.6 & $4.00\pm0.04$ & $100\pm12$ & $-176\pm12$ & $5.06\pm0.29$ & $149.8\pm1.7$ & & \\ 
\hfill LSB \dotfill & 220.6 & $3.91\pm0.03$ & $ 48\pm11$ & $-153\pm11$ & $4.08\pm0.28$ & $143.7\pm2.0$ & & \\ 
2005 Jun 6:\hfill\hfill & \nodata & \nodata & \nodata & \nodata & \nodata & \nodata & $-23.1\pm12.6$ & $ 251\pm  56$ \\ 
\hfill USB \dotfill & 348.0 & $4.22\pm0.04$ & $126\pm19$ & $-177\pm20$ & $5.13\pm0.47$ & $152.7\pm2.6$ & & \\ 
\hfill LSB \dotfill & 338.0 & $4.14\pm0.03$ & $ 94\pm16$ & $-214\pm15$ & $5.62\pm0.38$ & $146.8\pm1.9$ & & \\ 
2005 Jun 9:\hfill\hfill & \nodata & \nodata & \nodata & \nodata & \nodata & \nodata & $-5.0\pm1.7$ & $ 138\pm  17$ \\ 
\hfill USB \dotfill & 231.9 & $3.48\pm0.02$ & $-244\pm 9$ & $ -7\pm 8$ & $7.01\pm0.25$ & $90.8\pm1.0$ & & \\ 
\hfill LSB \dotfill & 221.9 & $3.38\pm0.02$ & $-224\pm 8$ & $ 28\pm 9$ & $6.68\pm0.24$ & $86.4\pm1.1$ & & \\ 
2005 Jun 15:\hfill\hfill & \nodata & \nodata & \nodata & \nodata & \nodata & \nodata & $-11.7\pm13.6$ & $ 192\pm  60$ \\ 
\hfill USB \dotfill & 348.0 & $3.31\pm0.02$ & $ 45\pm16$ & $-166\pm16$ & $5.18\pm0.48$ & $142.5\pm2.7$ & & \\ 
\hfill LSB \dotfill & 338.0 & $3.32\pm0.02$ & $ 29\pm14$ & $-181\pm13$ & $5.52\pm0.41$ & $139.6\pm2.2$ & & \\ 
2005 Jun 16:\hfill\hfill & \nodata & \nodata & \nodata & \nodata & \nodata & \nodata & $-5.4\pm1.8$ & $ 174\pm  18$ \\ 
\hfill USB \dotfill & 231.9 & $3.94\pm0.03$ & $-93\pm 9$ & $-196\pm 8$ & $5.50\pm0.22$ & $122.3\pm1.1$ & & \\ 
\hfill LSB \dotfill & 221.9 & $3.79\pm0.03$ & $-109\pm 8$ & $-157\pm 7$ & $5.05\pm0.19$ & $117.6\pm1.1$ & & \\ 
2005 Jun 17:\hfill\hfill & \nodata & \nodata & \nodata & \nodata & \nodata & \nodata & $-22.3\pm7.4$ & $ 246\pm  33$ \\ 
\hfill USB \dotfill & 348.0 & $2.95\pm0.02$ & $148\pm14$ & $-228\pm14$ & $9.21\pm0.48$ & $151.5\pm1.5$ & & \\ 
\hfill LSB \dotfill & 338.0 & $3.02\pm0.02$ & $109\pm12$ & $-276\pm12$ & $9.81\pm0.41$ & $145.8\pm1.2$ & & \\ 
2005 Jul 20:\hfill\hfill & \nodata & \nodata & \nodata & \nodata & \nodata & \nodata & $-7.5\pm1.6$ & $ 209\pm  16$ \\ 
\hfill USB \dotfill & 231.9 & $3.82\pm0.02$ & $ 15\pm 6$ & $-180\pm 6$ & $4.73\pm0.17$ & $137.3\pm1.0$ & & \\ 
\hfill LSB \dotfill & 221.9 & $3.75\pm0.02$ & $-25\pm 6$ & $-165\pm 6$ & $4.45\pm0.15$ & $130.7\pm1.0$ & & \\ 
2005 Jul 21:\hfill\hfill & \nodata & \nodata & \nodata & \nodata & \nodata & \nodata & $+1.1\pm8.2$ & $ 154\pm  36$ \\ 
\hfill USB \dotfill & 348.0 & $3.87\pm0.03$ & $240\pm18$ & $-220\pm17$ & $8.39\pm0.46$ & $158.7\pm1.6$ & & \\ 
\hfill LSB \dotfill & 338.0 & $3.73\pm0.03$ & $225\pm15$ & $-203\pm15$ & $8.11\pm0.40$ & $159.0\pm1.4$ & & \\ 
2005 Jul 22:\hfill\hfill & \nodata & \nodata & \nodata & \nodata & \nodata & \nodata & $-3.7\pm1.8$ & $ 152\pm  18$ \\ 
\hfill USB \dotfill & 231.9 & $3.37\pm0.02$ & $-91\pm 6$ & $-120\pm 6$ & $4.46\pm0.18$ & $116.4\pm1.1$ & & \\ 
\hfill LSB \dotfill & 221.9 & $3.34\pm0.02$ & $-105\pm 6$ & $-110\pm 6$ & $4.55\pm0.17$ & $113.1\pm1.1$ & & \\ 
2005 Jul 30:\hfill\hfill & \nodata & \nodata & \nodata & \nodata & \nodata & \nodata & $-4.8\pm1.4$ & $ 133\pm  14$ \\ 
\hfill USB \dotfill & 231.9 & $4.16\pm0.03$ & $-223\pm 7$ & $ 26\pm 6$ & $5.39\pm0.16$ & $86.6\pm0.8$ & & \\ 
\hfill LSB \dotfill & 221.9 & $4.12\pm0.03$ & $-193\pm 7$ & $ 53\pm 6$ & $4.86\pm0.15$ & $82.4\pm0.9$ & & \\ 
\enddata
\tablenotetext{a}{Statistical errors only. Overall flux density scale
uncertainty is 10\%.}
\end{deluxetable*}

To obtain full polarization information we used the SMA polarimeter
(\citetalias{MarroneE06}; \citealt{Marrone06}). Measurement of
polarization relies on precise determination of the fractional
contamination (``leakage'') of each polarization state by the
cross-handed polarization. Uncalibrated leakage contaminates the
linearly polarized Stokes parameters ($Q$ and $U$) with $I$. Leakages
were measured each night by observing polarized quasars (3C~279 or
3C~454.3) over a large range of parallactic angle. The leakages are
stable, approximately 1\% in the upper sideband (USB) for each band,
4\% for the 230~GHz lower sideband (LSB), and 2\% for the 345~GHz
LSB. The r.m.s. leakage variability is 0.3\% at 230 GHz and 0.4\% at
345~GHz. There are three potential sources of variability: (1) real
instrumental polarization changes between nights, (2) finite
signal-to-noise ratio, and (3) polarization leakage that is not
constant across the sky, resulting in leakage determinations that
depend on the hour angle coverage. The first of these may exist but is
bounded by the small observed leakage variability. The second is
expected to yield variations at the level of 0.1\% in the 345~GHz
leakages and less at 230~GHz. We know that the third effect is also
present, at a level of around 0.2\%, due to the fact that
cross-polarization introduced skyward of the Nasmyth relay mirror is
rotated relative to the feed as a source is tracked in elevation,
while the cross-polarization of the wave plate and subsequent optics
is fixed. This small additional leakage imprints itself on the
polarization of the calibrator and its effect on the leakages derived
in calibration (under the assumption of stationary leakage) changes
with the elevation coverage of the calibrator. Fortunately, the
resulting polarization changes are small. Leakage errors of
0.3\%$-$0.4\% contribute $0.2$\% fractional contamination to the
instantaneous polarization, significantly less when averaged over a
full track due to the parallactic angle rotation. For \sgra, which we
measure to be 5\%$-$10\% polarized, this results in at most 1\arcdeg\
of position angle error. We reduced the
\sgra\ data using the average leakages at each frequency instead of
the single-night values to confirm this estimate. More importantly for
this purpose, the uncorrected instrumental polarization is very nearly
constant across the sidebands because properties such as the
illumination pattern on the antenna change slowly with frequency. We
find the imprint of this additional leakage on the calibrator to be
very consistent between sidebands, so although the absolute position
angle varies by up to 1\arcdeg, the inter-sideband difference varies
only by 0.1\arcdeg$-$0.2\arcdeg. We take this as our systematic error
for position angle differences, although it is much lower than the
thermal noise. Previous SMA 345~GHz observations
\citepalias{MarroneE06} do not include enough calibration data to
place similar limits on the systematic errors, so we exclude them from
the following analysis.

\section{Rotation Measure and Intrinsic Polarization}
\label{s-RMX}
\begin{figure}
\plotone{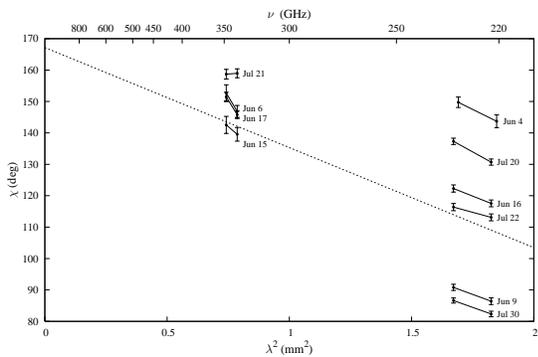}
\caption{The position angles observed in each sideband on each of the
ten epochs. The slope between the two sidebands at each epoch is
proportional to the RM and the extrapolated intercept at the $\chi$
axis is $\chi_0$. The mean RM$-$$\chi_0$ fit is also
plotted. Variability in $\chi_0$ is visible in the 230~GHz points (at
right), which show slopes similar to the mean but are widely dispersed
in $\chi$.}
\label{f-RMplot}
\end{figure}

Faraday rotation changes the observed polarization position angle
($\chi$) as a function of frequency according to:
\begin{equation}
\chi(\nu) = \chi_0 + \frac{c^2}{\nu^2}\mathrm{RM} ,
\label{e-RM}
\end{equation}
where $\chi_0$ is the intrinsic position angle. The rotation measure
(RM) is proportional to the integral of the electron density and
magnetic field component along the line of sight. From the observed
LSB and USB position angles we derive a RM and $\chi_0$ for each
observation. We plot $\chi$ in both sidebands against $\lambda^2$ for
all ten epochs in Figure~\ref{f-RMplot}, along with the average RM and
$\chi_0$. The larger errors in Table~\ref{t-polObs} for 345~GHz band
RMs and $\chi_0$s are due to the much smaller inter-sideband
difference in $\lambda^2$ at the higher frequency; the constraints on
these quantities are therefore dominated by the 230~GHz data. The
average RM from all 10 epochs is $(-5.6\pm0.7)\times10^5$~\rmm, while
the 230~GHz and 345~GHz points alone yield
$(-5.4\pm0.7)\times10^5$~\rmm\ and $(-13\pm5)\times10^5$~\rmm,
respectively, consistent within their errors. The ten single-night RM
values are consistent with a constant value ($\chi^2_r = 1.21$, for 9
degrees of freedom). None of the 230~GHz points deviates from the
average by more than 1.2 times its measurement error, while the
largest deviation at 340~GHz is 2.2$\sigma$. We therefore place a
conservative upper limit of three times the averaged $\sigma$, or
$2\times10^5$~\rmm\ on RM variability. This agrees with the small RM
fluctuations inferred from the constancy of polarization at 83~GHz
\citep{MacquartE06}.

These observations also constrain the intrinsic polarization direction
of \sgra. The average $\chi_0$ in our ten measurements is
$167^\circ\pm7^\circ$, or $162^\circ\pm7^\circ$ and
$210^\circ\pm21^\circ$ from 230 and 345~GHz observations,
respectively. The $\chi_0$ values vary by more than our measurement
errors predict, suggesting intrinsic polarization changes. Assuming a
constant $\chi_0$, we obtain $\chi^2_r=2.8$ (0.3\% probability) using
all data points, or $\chi^2_r=3.9$ (0.16\%) for the 230~GHz points
only. The scatter in the full data set suggests an intrinsic $\chi_0$
dispersion of $31^{+18}_{-9}$ degrees, with very similar results
obtained from just the 230~GHz points. It is interesting to note that
the mean $\chi_0$ is $\sim$80\arcdeg\ different from the angle
observed during IR flares \citep{EckartE06-lp,MeyerE06}. A 90\arcdeg\
position angle change has been predicted to occur near the spectral
peak in theoretical models of the polarization, although the precise
flip frequency depends on the details of the models
\citep{Agol00,MeliaLiuCoker00}. If the IR measurements trace the
intrinsic polarization direction at short wavelengths, rather than the
possibly random polarization of the transient flare, this would be
evidence for such a change above 345~GHz. We do not observe a flip
between 230 and 345~GHz.

This measurement represents the first reliable determination of the RM
of \sgra, and the only measurement made from simultaneous observations
at multiple frequencies and therefore able to isolate source
polarization changes. \citet{MacquartE06} derived a RM of
$-4.4\times10^5$~\rmm\ from the average $\chi$ at four frequencies
over the last several years. However, their interpretation requires a
180\arcdeg\ unwrapping of the 83~GHz position angle and the
non-wrapped position angle is not strongly excluded. They report a
$\chi^2_r$ of 2.1 for 3 degrees of freedom (10\% probability) for the
non-wrapped fit (RM=$-1.9\times10^5$~\rmm). Furthermore, this
$\chi^2_r$ relies on their poorly measured USB polarization (just two
of five measurements detect polarization at $3\sigma$) and the
standard error of the mean of just two measurements at 216~GHz in
order to provide additional degrees of freedom. Discarding the former
and using the 230~GHz variability to estimate the 216~GHz variability
increases the probability of the non-wrapped position angle to 26\%.

\section{Discussion}
\label{s-disc}
The RM we observe is too large to be produced by material beyond the
accretion radius of \sgra\ (approximately 1\arcsec\ or 0.04~pc). Using
the density determined by \citet{BaganoffE03}, the RM in the dense
inner 10\arcsec\ is just $8\times10^3$~\rmm\ assuming a 1~mG ambient
field. RMs determined in nearby sources are as large as 70\% of this
estimate \citep[e.g.][]{YZMorris87}.

The RM can be used to determine the accretion rate ($\dot M$) at small
radii around \sgra\ if assumptions are made about the nature of the
accretion flow. The procedure is outlined in \citetalias{MarroneE06}
and assumes a power-law radial density profile ($n\propto r^{-\beta}$)
and an ordered, radial, equipartition-strength magnetic field. Our RM
detection places upper limits that are 15\% lower than those in their
Figure~4. The important parameter for determining the upper limit is
\rin, the radius at which electrons become relativistic and their RM
contribution begins to be suppressed \citep{QuatGruz00-LP}. In
general, shallow density profiles approaching that of the
convection-dominated accretion flow ($\beta\rightarrow 1/2$;
\citealt{NarayanE00,QuatGruz00-CDAF}) yield hotter central
temperatures and thus larger \rin\ \citep{Quataert03}, while steep
profiles ($\beta\rightarrow3/2$), as in Bondi-like or
advection-dominated accretion flows (ADAF; \citealt{NaraYi95}) are
marginally relativistic to small radii and may have \rin\ of a few
$r_\mathrm{S}$. Simulations
\citep[e.g.][]{HawleyBalbus02,IgumenshchevE03,PenE03} favor
$\beta\le1$. Those with published temperature profiles show \rin\
between 30 and 100 Schwarzschild radii ($r_\mathrm{S}$), yielding
upper limits for $\dot M$ of $2\times10^{-7}$ to
$5\times10^{-8}$~\msunyr.

Note that the assumptions of the \citetalias{MarroneE06} formalism,
which were taken from previous related works
\citep[e.g.][]{Melia92,QuatGruz00-LP}, are not constrained by existing
observations. In particular, the assumption of equipartition-strength
fields cannot be justified observationally, although simulations may
show a tendency to reach equipartition fractions of a few percent
\citep{IgumenshchevE03}. Magnetic fields that are a fraction
$\epsilon$ of the equipartition strength will raise the accretion rate
limits by $\epsilon^{-2/3}$ (a factor of 10 for $\epsilon=3$\%).  The
assumption of a non-reversed field is also potentially suspect. The
direction of position angle change with frequency, and thus the sign
of the RM, appears unchanged since the \citet{AitkenE00} measurements
in 1999. This is not natural for a turbulent accretion flow, except
along special lines of sight (e.g., the axial region in model A of
\citealt{IgumenshchevE03}). \citet{RuszBegel02} suggested that the
stability of the sign of circular polarization (CP) noted by
\citet{BowerE02} could be explained by Faraday conversion of linear to
circular polarization in a highly reversed field with a small
directional bias. The bias field ($B_b$) naturally creates stability
in the CP and RM signs if $B_b \gg B_{rms}/\sqrt{N}$, where $B_{rms}$
is the random field and $N$ the number of reversals. In this limit the
integrated product of the electron density and parallel magnetic
field, without regard to field reversals, would be large, as would the
accretion rate, but the net RM would be small and originate in the
bias field. The turbulent jet model of \citet{BeckertFalcke02} also
uses a random field with a small bias to produce stability. If these
models are correct the accretion rate upper limits derived from our RM
detection would no longer hold.

Our RM also allows us to place lower limits on the accretion rate;
these are not subject to the above caveats since the uncertainties act
to raise the minimum accretion rate. If we take \rin\ to be around
$10r_\mathrm{S}$ or $3r_\mathrm{S}$ (smaller \rin\ yields smaller
lower limits, so this is conservative for hot flows), we find that
$\dot M$ must be greater than $1-2\times10^{-8}$~\msunyr\ or
$2-4\times10^{-9}$~\msunyr, respectively. If the field is toroidal,
reversed, or sub-equipartition, these lower limits are raised and may
pose problems for very low-$\dot M$ models. Using the RM-$\dot
M^{3/2}$ scaling derived in \citetalias{MarroneE06}, the observed
limit on RM variability (35\%) limits accretion rate variability to
25\% over two months.

Our detection of significant variability in $\chi_0$ is the first
clear separation of the effects of a variable RM and variable
intrinsic polarization; disentangling these effects requires
multi-frequency observations. In the optically thin limit $\chi_0$
represents the intrinsic polarization direction of \sgra\ at all
frequencies and $\chi$ variations should follow $\chi_0$ variations,
with RM fluctuations potentially contributing a frequency-dependent
component to the $\chi$ dispersion. However, comparing all previous
measurements near 230 and 345~GHz (employing our RM to translate all
observations to a single frequency), we find a dispersion of
22\arcdeg\ in 16 observations at 230~GHz and just 8\arcdeg\ in 11
epochs at 345~GHz (\citealt{AitkenE00,BowerE03,BowerE05};
\citetalias{MarroneE06}; this work). Although these two dispersions
scale by approximately $\nu^{-2}$, consistent with RM fluctuations of
approximately $2\times10^5$~\rmm, we consider this explanation
unlikely. From equation~(\ref{e-RM}), the proportionality between RM
and $\chi$ fluctuations should be $\lambda^2$ if RM changes are
responsible for $\chi$ changes, with an increasing RM (less negative)
expected for increasing $\chi$. In fact our 230~GHz data show the
opposite trend, with the least negative RM values occurring on the days
with the smallest $\chi$. The best-fit slope of the RM$-$$\chi_{230}$
relation is $(-3.4\pm3.6)\times10^3$~rad~m$^{-2}$~deg$^{-1}$,
different from the expected slope of
$+10\times10^3$~rad~m$^{-2}$~deg$^{-1}$ by $4\sigma$. Similar analysis
is inconclusive in the 345~GHz data, which are less sensitive to
RM. This confirms that we are seeing intrinsic polarization changes
rather than RM fluctuations. Moreover, the necessary RM changes would
yield variability of 150\arcdeg\ at 83~GHz, much larger than the
8\arcdeg\ observed by \citet{MacquartE06}. The emission from \sgra\
appears to be optically thick below $300-400$~GHz (Marrone et al., in
prep) so $\chi_0$ may vary with frequency, contrary to the expectation
at low optical depth. The observed spectrum of position angle
variability between 83 and 345~GHz most likely represents differences
in the stability of the magnetic field orientation (manifested as
$\chi_0$) at the photospheres at these frequencies. If the Faraday
screen is located at much larger radii than the submillimeter emission
region, its variations are likely to be slower than those of the
intrinsic polarization; this allows the intrinsic changes to be
isolated from putative RM fluctuations by examining short timescales,
even without better knowledge of the RM stability.

\acknowledgements
The authors thank the SMA staff, particularly Ken Young, for
assistance with the polarimeter, and Avi Loeb and Ramesh Narayan for
useful discussions. DPM acknowledges partial support from a Harvard
University Merit Fellowship.


\end{document}